*Laser initiated p-$^{11}$B fusion reactions in petawatt high-repetition-rates laser facilities*

M. Scisciò[1*], G. Petringa[2], Z. Zhu[3,4,5], M. R. D. Rodrigues[3], M. Alonzo[1], P. L. Andreoli[1], F. Filippi[1], Fe. Consoli[2], M. Huault[6], D. Raffestin[7], D. Molloy[8], H. Larreur[6,7], D. Singappuli[7], T. Carriere[7], C. Verona[9], P. Nicolai[7], A. McNamee[8], M. Ehret[10], E. Filippov[10], R. Lera[10], J. A. Pérez-Hernández[10], S. Agarwal[11], M. Krupka[11,12], S. Singh[11,12], V. Istokskaia[21], D. Lattuada[2,13], M. La Cognata[2], G. L. Guardo[2], S. Palmerini[14,15], G. Rapisarda[2], K. Batani[16], M. Cipriani[1], G. Cristofari[1], E. Di Ferdinando[1], G. Di Giorgio[1], R. De Angelis[1], D. Giulietti[17], J. Xu[4,18], L. Volpe[10,19], M. D. Rodríguez-Frías[10,20], L. Giuffrida[2,21], D. Margarone[2,8,21], D. Batani[7], G. A. P. Cirrone[2], A. Bonasera[3] and Fa. Consoli[1*]

1 ENEA Nuclear Department - C. R. Frascati, Via Enrico Fermi 45, 00044 Frascati, Italy

2 Laboratori Nazionali del Sud, Istituto Nazionale di Fisica Nucleare (LNS-INFN), Catania 95125, Italy

3 Cyclotron Institute, Texas A&M University, College Station, TX 77843, USA

4 Shanghai Institute of Applied Physics, Chinese Academy of Sciences, Shanghai 201800, China

5 University of Chinese Academy of Sciences, Beijing 100049, China

6 Departamento de fisica fundamental, Universidad de Salamanca, Patio de Escuelas 1, 37008 Salamanca, Spain

7 Université de Bordeaux, CNRS, CEA, CELIA (Centre Lasers Intenses et Applications), Unité Mixte de Recherche 5107, Talence 33400, France

8 Queen's University Belfast, School of Mathematics and Physics, Belfast BT7 1NN, UK

9 Dipartimento di Ingegneria Industriale, Università di Roma "Tor Vergata", Via del Politecnico 1, 00133, Roma, Italy

10 Centro de Laseres Pulsados, Building M5, Science Park, Calle Adaja 8, Villamayor, 37185 Salamanca, Spain

11 FZU-Institute of Physics of Czech Academy of Sciences, 182 21 Prague, Czech Republic

12 Institute of Plasma Physics of Czech Academy of Sciences, 182 00 Prague, Czech Republic

13 Facoltà di Ingegneria e Architettura, Università degli Studi di Enna "Kore", 94100 Enna, Italy

14 Dipartimento di Fisica e Geologia, Università degli Studi di Perugia, Via A. Pascoli s/n, 06123 Perugia, Italy

15 Istituto Nazionale di Fisica Nucleare, sezione di Perugia, Via A. Pascoli s/n, 06123 Perugia, Italy

16 IPPLM Institute of Plasma Physics and Laser Microfusion, Hery Street 23, 01-497, Warsaw, Poland

17 Dipartimento di Fisica, Università di Pisa and INFN, Largo B. Pontecorvo, n.3, 56127 Pisa, Italy

18 School of Physics Science and Engineering, Tongji University, Shanghai 200092, China

19 ETSI Aeron.utica y del Espacio, Universidad Politécnica de Madrid 28006, Madrid, Spain

20 Departamento de F.sica y Matem.ticas, University of Alcal., Plaza de San Diego s/n, 28801 Madrid, Spain

21 ELI Beamlines Facility, The Extreme Light Infrastructure ERIC, Dolni Brezany 252 41, Czech Republic

* Massimiliano.sciscio@enea.it , fabrizio.consoli@enea.it

## Abstract

Driving the nuclear fusion reaction p+11B→ 3α + 8.7 MeV in laboratory conditions, by interaction between high-power laser pulses and matter, has become a popular field of research, due to numerous applications that it can potentially allow: an alternative to deuterium-tritium (DT) for fusion energy production, astrophysics studies and alpha-particle generation for medical treatments. A possible scheme for laser-driven p-$^{11}$B reactions is to direct a beam of laser-accelerated protons onto a boron sample (the so-called "pitcher-catcher" scheme). This technique was successfully implemented on large, energetic lasers, yielding hundreds of joules per shot at low repetition. We present here a complementary approach, exploiting the high-repetition rate of the VEGA III petawatt laser at CLPU (Spain), aiming at accumulating results from many interactions at much lower energy, for better controlling the parameters and the statistics of the measurements.

Despite a moderate energy per pulse, our experiment allowed exploring the laser-driven fusion process with tens (up to hundreds) of laser shots. The experiment provided a clear signature of the produced reactions and of the fusion products, accumulated over many shots, leading to an improved optimization of the diagnostic for these experimental campaigns In this paper we discuss the effectiveness of the laser-driven p-$^{11}$B fusion in the pitcher-catcher scheme, at high-repetition rate, addressing the challenges of this experimental scheme and highlighting its critical aspects. Our proposed methodologies allow evaluating the performance of this scheme for laser-driven alpha particle production and can be adapted to high-repetition rate laser facilities with higher energy and intensity.

1.  **Introduction**

The interaction of intense laser pulses with matter has been historically exploited for triggering nuclear fusion reactions. Deuterium-tritium (DT) fuel is the best-known candidate for future reactors for energy production, due to the lowest temperature required to initiate the fusion process [1]. This reaction requires radioactive combustible (T), of extremely scarce availability, and generates also energetic neutrons, besides alpha particles, that induce activation on the materials they interact with, making the setting and the maintenance of fusion reactors a serious issue. Moreover, the energy conversion efficiency of neutrons is rather low. These characteristics of DT fuel have encouraged research activity on alternative fuels. Exploiting the p-$^{11}$B reaction is to date one of the most appealing solutions, since energy is released in terms of alpha particle kinetic energy. Both reagents are abundant in nature and not radioactive, and, at high energies (> 500 keV), the p-$^{11}$B and DT cross sections are comparable. This motivates the increasing interest that p-$^{11}$B fusion achieved in recent years, by both public institutions and private companies, for unveiling the actual applicability for schemes of energy production [2], [3], [4], and also for potential use for astrophysics [5], medical treatments [6] and applications related to localized sources of alpha particles.

The first demonstration of a laser-driven p-$^{11}$B reaction used a picosecond laser pulse at intensity about $10^{18}$ W/cm$^2$, directly focused onto a composite target $^{11}$B+(CH$_2$)$_n$, resulting in about than $10^5$ detected alpha particles in 4π steradians [7], [8], followed by similar yield on experiments with $10^{15}$ W/cm$^2$ intensity with ABC ns-laser [5]. The use of far more energetic lasers (about 500 J) with ~300 ps pulse of Full-Width-Half-Maximum (FWHM), about $10^{16}$ W/cm$^2$ on advanced targets (H-B-enriched solid Si, or BN containing H due to the target synthesis) led to the remarkable value of about $10^9$-$10^{10}$ alpha particles per steradian [9], [10], [11], far higher than what achieved in the past, giving a strong impulse to the research [2], [3], [12], [13], [14], [15], [16].

An alternative scheme consists of directing protons accelerated by laser-matter interaction, either on solid B targets or on B plasmas created by a nanosecond-pulse laser on B solid target [16], [17], [18], [19], with demonstration of about $10^7$ alpha particles per steradian. This 'pitcher-catcher' scheme, recently applied on the high-energy LFEX laser (Japan) on BN secondary target, led to about 5 ×$10^9$ alpha particles per steradian [20].

Both the aforementioned schemes (in-target and pitcher-catcher) were exploited with energetic lasers - with hundreds to thousands of joules per shot. This is possible on a few large installations but limited to very few shots. Moreover, the wide and intense secondary radiation (X, gamma, electrons, ions), produced with a single energetic shot, gives significant limitations to the experiment diagnostics. Instead, we propose in this work to use a complementary approach, assessing and exploiting the effectiveness of high-repetition rate experiments in pitcher-catcher configuration for petawatt-scale laser-triggered p-$^{11}$B fusions. The accumulation of results from a large number of interactions with lower laser pulse energy and at 'moderate' intensity, may provide a better control of the parameters and of the statistics of the measurements. Previous experimental studies were performed on this purpose, such as on the high repetition rate ECLIPSE laser [12], [14], [21] (35 fs FWHM, 110 mJ, $2\times10^{18}$ W/cm$^2$), or using kHz repetition rate at GW power [22].

We present in this paper results achieved using an upgraded setup on the VEGA III laser at CLPU in Spain. The laser has characteristics similar to ECLIPSE but with significantly higher energy per pulse. Our aim was to develop, test and compare new experimental schemes for studying and exploiting p-$^{11}$B reactions in high-power, high-repetition rate lasers. The main objective of the campaign was to assess the issues of the scheme, to improve alpha production efficiency and ensure accurate detection of p-$^{11}$B reactions triggered by laser-matter interactions. Therefore, we conducted a precise characterization of the interactions and employed advanced techniques for alpha yield estimation [13].

## 2. **Experimental setup and diagnostics**

The experimental setup employing the pitcher-catcher scheme to induce laser-driven p-$^{11}$B fusion reactions, is qualitatively illustrated in Fig. 1(a) and (b). Since the p-$^{11}$B reaction cross-section exhibits two resonances at 148 keV and 614 keV center-of-mass energy [17] and maintains high values up to several MeVs [23], we aimed at accelerating a substantial flux of TNSA protons from thin-foil targets, for increasing the number of protons with energies up to a few MeV. Since we did not aim at very high energies, laser focal intensities up to ~4 × 10$^{19}$ W/cm$^2$ were considered. With such moderate intensities – for the characteristics of VEGA III (nominally, 30 J maximum pulse energy, 30 fs minimum pulse duration, 800 nm wavelength, contrast $2\times10^{-5}$ at 1 ps before the main pulse and <10$^{-5}$ at 5 ps) - it was possible to relax the laser parameters in order to improve the shot-to-shot repeatability. This required an experimental optimization, leading to the use of long focal lengths optics, laser pulses of average duration ~220 fs, ~10 μm focal spot diameter on target and a laser pulse energy (in average, during our campaign) of 27 J before the pulse compression with about 25% energy deposition on target. These conditions, not very usual for petawatt femtosecond lasers, required a detailed characterization of the interaction and of the emitted particle flows, with the purpose to provide a picture as thorough as possible, for enabling effective coupling with the theoretical models that will be discussed in Section 8 and in future works in preparation [24]. As a result of this optimization, the pitcher target was set to a 6 μm thick aluminum foil. It was irradiated by the laser with an angle of 12° with respect to its normal, generating the laser-accelerated protons that impinged the boron catcher. This secondary target was 2 mm thick and had a surface of 27 x 27 mm and was carefully positioned to maximize the number of TNSA ions impinging on it. The proton-irradiated boron catcher, where p-$^{11}$B fusion reactions took place, ideally emits three alpha particles for each fusion reaction and momentum conservation entails that higher energies are expected in the direction of the incoming proton beam and lower in the opposite direction. This will play a role with increased importance for protons with higher energies. However, we are interested on alphas capable to escape the B bulk, so those originated at a maximum depth of the order of a few tens of microns from the B front surface. For instance, alpha particles with 5 MeV (i.e. the peak energy of the theoretically expected alfa spectrum at the main 660 keV resonance) are stopped by only 18 μm of B and will therefore remain inside the sample if they are generated inside the bulk. Hence, protons up to a few MeVs are of interest [13], and because of the much larger mass of B, this effect of downshifted alpha spectrum should be not extremely large. These considerations motivated the decision to tilt the B at the reasonably large (54° degree) angle relatively to the pitcher normal, for enhancing the number of alpha particles to be revealed by the CR39 diagnostics (see Fig. 1(b)) after escaping the Boron. The tilting is meant in fact to increase the component of the proton momentum tangential to the B surface, and consequently to reduce the one along its normal. In this way, for the same proton energy, the proton stopping position is closer to the B surface, and alphas produced from p-$^{11}$B reactions will need to pass through a shorter path in B to exit from the bulk. The choice of the angle also considered the necessity of not extending the interaction region too much, on one side, and the improve the feasibility of monitoring of the fusion products stemming out of the boron, on the other side. As reported in Fig. 1(b), the tilted B catcher had a distance of ~13 mm from the Al pitcher. The catcher was placed on a motorized stage, to be displaced during the laser shots when the TNSA proton beam was characterized, with the diagnostics placed along the target normal (which would have been blinded by the boron sample). This allowed us to dedicate part of the campaign to the careful characterization of the TNSA particles.

The setup shown in Fig. 1(a) includes diagnostics for both the characterization of the TNSA proton beam from the Al target and the revelation of the fusion products stemming out of the proton-irradiated boron sample. The cartesian coordinate system of Fig. 1(a) serves here as a reference for the equatorial angle $\theta$ on the xy-plane (i.e. the plane of the interaction point), being $\theta = 0$ the laser axis. On the Al target normal, i.e. at $\theta = 12°$, at distance ~720 mm from Target Chamber Center (TCC) a Thomson spectrometer (indicated with "TS 1" in Fig. 1(a)), equipped with a MCP detector, allowed measurements of the energy spectra of the laser-generated protons and heavy ions, at a high repetition rate [25]. During the campaign, we also implemented multiple diamond-based Time-of-Flight detectors (not shown in the condensed schemes of Fig. 1), placed at different angles with respect to the Al target, for obtaining a full view of the angular distribution of the TNSA ions. This analysis, too extensive for being included in this work, is discussed in detail in Ref.[24], The laser interaction with the Al target was monitored also with electron spectrometers placed on the target equatorial plane, to characterize the TNSA mechanism accelerating ions. The first in laser-forward direction, with respect to the target, at an angle $\theta = 21°$ (i.e. at 9° from target normal) at a distance of 668 mm from TCC. The second, in laser-backward direction, at an angle $\theta = -139°$ (i.e. at 41° from target normal) at a distance of 399 mm.

For revealing the alpha particles produced by the p-$^{11}$B reactions, which had sufficient energy to escape the B catcher, we implemented an array of CR39 detectors [26], horizontally aligned as reported in Fig. 1(b), at 250 mm from the boron and in line with the interaction point of the proton beam with the B sample, aiming at covering part of the wide cone of emission from it. The array was ~100 mm long, featured three detectors of dimensions 20 x 20 mm, equally spaced at three positions (indicated in Fig. 1(b) with P1, P2 and P3), and divided into regions with different filter thicknesses (as will be discussed more into detail in the Section 7). With respect to the horizontal plane including the point of interaction, the array was placed about 20 mm below. These detectors were exposed to multiple shots series, for accumulating the alpha reaction products of several proton-boron interactions. In a similar position of the CR39, at a distance of ~250 mm from the boron target, we placed an additional Thomson spectrometer (TS 2) [27], to support CR39 measurements, exploiting the capability of the TS to discriminate protons from heavy ions. The TS 2 was equipped with either CR39 detector or Imaging Plate detectors, depending on the shot series, and accumulated signals of particles stemming out of the boron sample during several laser shots. As detailed in Appendix B, the implementation of the aluminum shield (2 mm thick) shown in Fig. 2(b) was of key importance to protect the CR39 detectors and the TS 2 from the TNSA ions emitted at a large angle that would directly impinge the detectors and enter the TS 2.

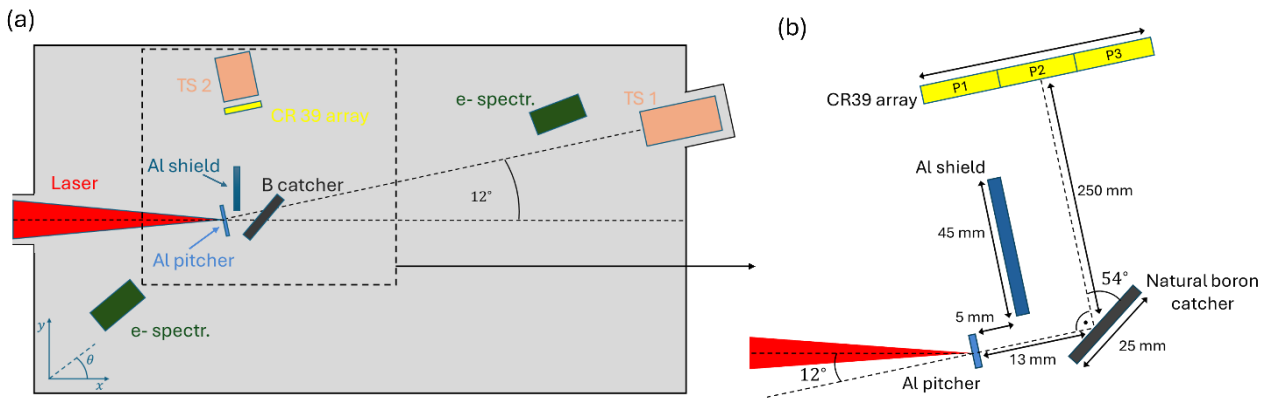

*Fig. 1 – (a) Top view of the experimental setup. The laser impinges on the Al "pitcher" target (6 μm thickness) with a 12° angle. Ion and electron diagnostics are shown, together with the Al shield. . (b) Zoomed view of the pitcher-catcher configuration and the CR39 detector array (the pictured distances, dimensions and angles are not in scale to each other). The horizontal width of P1+P2+P3 is 100 mm.*

Due to the implemented pitcher-catcher scheme, exploiting TNSA protons that have energies from the keV range to the multiple MeV range, it is expected that many of the alphas will remain within the catcher, as discussed in reference [13]. They will be produced inside the catcher's bulk, due to the high penetration of the protons, and will not have enough energy to escape the B sample. For this reason, for evaluating their overall number, we relied upon a methodology different from CR39s. The natural B, used as catcher, contains about 20% of $^{10}$B isotopes and 80% $^{11}$B isotopes. In this way, the TNSA proton beam simultaneously interacted with both $^{11}$B and $^{10}$B, driving, among others, the reaction p+$^{10}$B → α + $^{7}$Be, producing radioactive $^{7}$Be isotopes with a half-time decay of about 53 days. The number of produced $^{7}$Be isotopes was measured on-site with a high purity Ge-detector (HPGe), Canberra XtRa Model GX301932 [28] used for the γ-spectroscopic analysis of its 478 keV peak. It was equipped with the CAEN DT5781 quadruple independent 16 k digital MCA and CoMPASS acquisition software [29] that permitted to supply a timestamp for each registered event. Further details on these measurements and on the detector and setup calibrations can be found in reference [28] of the same authors. This system allowed us to get off-line information on the effectiveness of interaction, cumulating over many shots [28]. The long decay time of $^{7}$Be with respect to shot-to-shot delay in VEGA III, allowed observing a cumulative effect on the B. Starting from this, the characterization of the TNSA proton beam in terms of particle flux allowed us, from the number of measured $^{7}$Be isotopes, estimating the number of induced p-$^{11}$B fusion reactions occurring simultaneously on the same B and, therefore, the number of alpha particles produced. The natural B catcher can be also radioactivated due to p+B → $^{11}$C + n - 2.9 MeV, producing the $^{11}$C isotope, of important interest for medical applications [28], [30]. The $^{11}$C has half-life of about 20 minutes and gamma peak at 511 keV, that can be useful to get information on the number of produced alpha particles from the p-$^{11}$B reactions for energies larger than 3 MeV as discussed in Ref. [28].

3. **Characterization of the accelerated TNSA beams**

The Thomson spectrometer (TS 1) was placed on the pitcher target normal (see Fig. 1(a)) and served as the main diagnostic for estimating the flux and the maximum energy of the laser-generated protons. These results were the main output used for optimizing, during the first part of the campaign, the laser features in terms of pulse duration and focal position, and the pitcher parameters in terms of target material and thickness. The B catcher was removed from the setup for allowing a line-of-sight of the TS to the Al target. In Fig. 2, we report typical proton and carbon ions spectra (measured by the TS 1), obtained from 6 μm Al targets, where the laser pulse energy was, on average, ∼ 7 J energy on target, and the mean pulse duration was ∼220 fs. These laser conditions were obtained after a first phase of optimization and resulted in producing the best proton beam (in terms of particle flux and cut-off energy). The spectrum of carbon ions is obtained by adding all measured ionization states coming from the relative ion parabolas of the same TS 1, which were singularly analysed (the high energy end of the distribution is mainly due to C5+ and C+6 ions). It is assumed that the entire contribution on these parabolas is only due to carbon ions, and the produced spectra are useful for the estimations discussed later in the paper. This set of laser parameters was therefore utilized during the second phase of the experiment where multiple shot series were used for driving p-$^{11}$B reactions and accumulating the fusion products with the dedicated diagnostics. However, despite the optimization process for achieving here the best stable beam conditions, the reported error-bars (which represent the standard deviation of the particle flux among the shots used for the average) indicate significant shot to shot fluctuations occurred during the entire experiment.

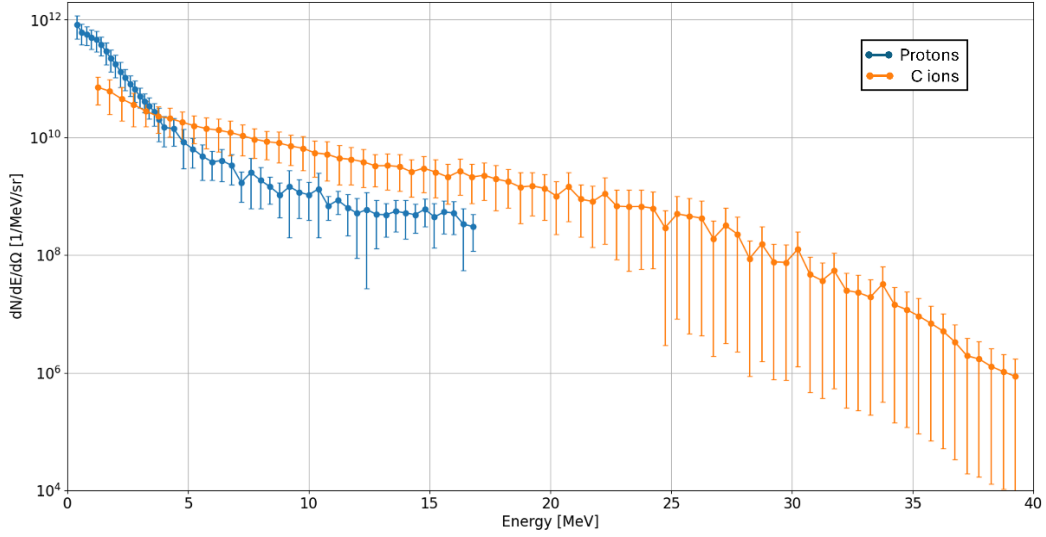

*Fig. 2 – Typical spectrum of the protons (blue) and carbon ions (orange) measured at the Al target normal. The spectra were obtained by averaging the data from 31 shots (protons) and 19 shots (carbon ions); the error-bars indicate one std. deviation.*

### 4. **Analysis of laser plasma interaction: electron spectra**

Additional information about the laser-plasma interaction occurring in the primary target was given by the measurements performed with the electron spectrometers [31]. On the rear side of the Al target, a spectrometer, optimized for energies up to 55 MeV, was placed at 9° from target normal. On the front side of the Al target, where the laser impinged, we placed a spectrometer for electrons up to 35 MeV, at 41° from target normal. The spectra obtained from these two devices are reported in Fig. 3(a), where the top plot represents the rear side and the bottom plot the front side. These energy spectra were obtained by accumulation of 123 shots on the Imaging Plate detectors inside the spectrometers and are reported here as an average of these shots. They are a representation of a routinely obtained electron spectrum, from a single laser shot. For estimating the relativistic temperatures of the hot and cold component of the accelerated electrons, we fitted the high-energy spectra with a Maxwell-Juttner distribution [32]:

$$f(\gamma) = \frac{\gamma\sqrt{\gamma-1}}{\theta K_2(1/\theta)} e^{-\gamma/\theta}$$

Being $\gamma$ the electrons' relativistic factor, $\theta = k_B T/m_e c^2$ the electron temperature and $K_2$ the modified Bessel function of second kind. The fit obtained for the spectrum on the target front side is reported in Fig. 3(b). We obtained temperatures $k_B T = 0.511\ MeV$ and $k_B T = 3.125\ MeV$ for the cold and hot component, respectively. These temperatures are similar to those obtained from the proton spectra [28], showing the common source mechanism [33].

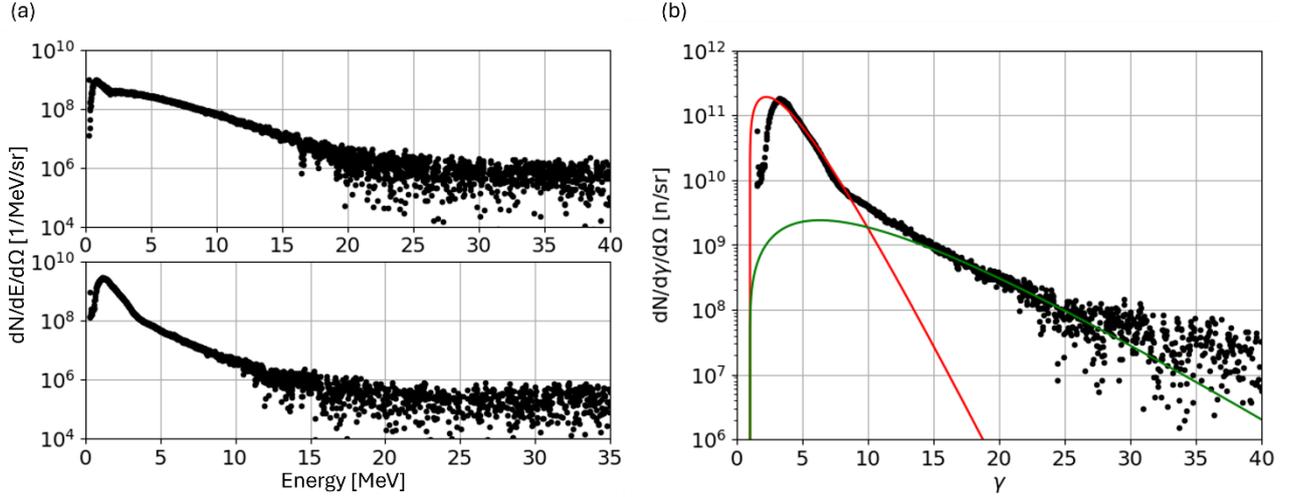

*Fig. 3 – (a) Electron spectrum measured on the Al target rear side (top plot) and front side at (bottom plot). Both spectrometers were placed on the target equatorial plane. The spectra are obtained from averaging over 123 laser shots. (b) Electron spectrum from the front side of the Al target, fitted with two Maxwell-Juttner functions, for the cold and hot component of the laser-accelerated electrons. The obtained temperatures are 0.511 MeV and 3.125 MeV.*

5. **Characterization of the irradiated Boron sample**

During the experiment, different nominally-identical B samples were used as catcher targets and were irradiated by multiple-shot series. The accumulation of many shots onto the same B left, on the irradiated surface, an area where debris coming from the pitcher (mainly Al) is deposited after being sputtered by the laser hitting the target. This gave information on aging of the B surface after many shots, on one side and, on the other side a visual estimation on where the main irradiation of TNSA beam on B surface occurred, and then where alpha particles, produced by p-$^{11}$B fusion reactions, were emitted from. In Fig 4(a) we report a microscope image of the irradiated boron. The horizontal and vertical lineouts of the gray values of the image (indicated in the figure with the red dashed lines), provide a good estimation of the elliptic area that is marked by the debris. We fitted the grey values of the image, plotted in Fig. 4(b) and 4(c) as a function of the width and height, with Gaussian-like functions that provided information about the dimension of the marked area. On the horizontal lineout we estimated the width of the "marked" area with the sigma of the fit-function, i.e. $\sigma = 2.8$ mm. On the vertical lineout, the "marked" area extends over the bottom of the sample (as one can recognize from the drop of the grey value, on the vertical lineout, at about 3 mm). The dimension can still be evaluated by the fit function, which yields $\sigma = 4.8$ mm. Both vertical and horizontal lineouts indicate that the peak of the grey values is on the irradiated target, indicating that the sample was well aligned with the Al target. The marked area has an elliptic shape, as expected, since the boron sample had an inclination of 54° with respect to the pitcher. Using the obtained values of horizontal and vertical $\sigma$ as the major and minor axis of the elliptic marked surface, we obtain an area $A \approx 42.2\ mm^2$.

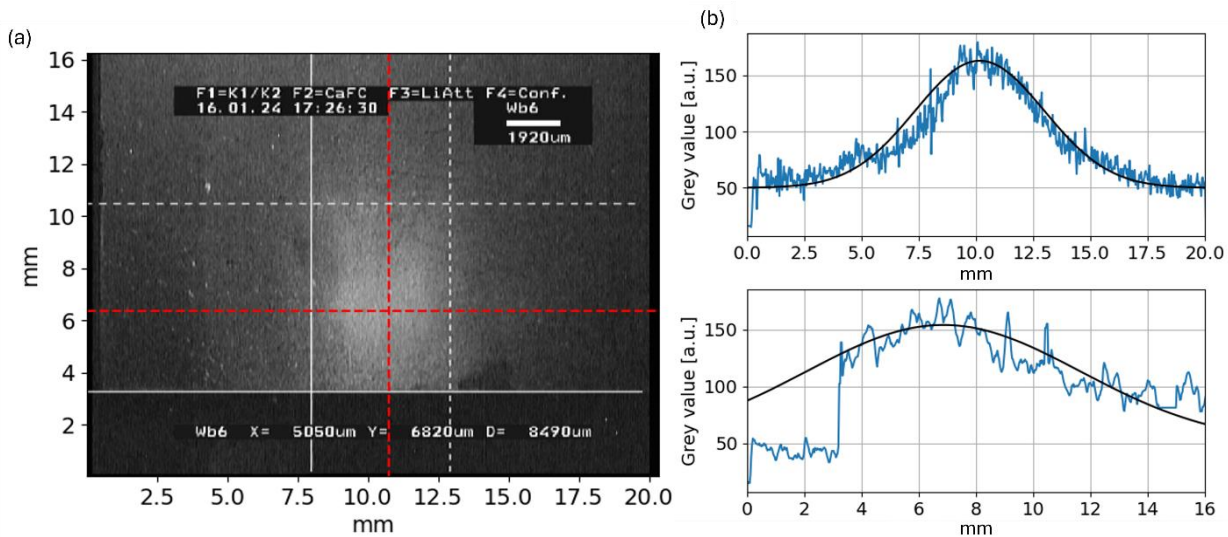

*Fig. 4 – (a) Microscope image of the irradiated boron sample, after multiple shot irradiation. The "marked" areas indicate the presence of aluminium debris on the irradiated surface and can serve as an indication on the cone of emission of the TNSA protons. (b) Lineouts (horizontal on top and vertical on bottom) of the grey values of the microscope image (indicated on the image on panel (a) by the red dashed lines).*

## 6. Thomson spectrometer pointing onto the B catcher

The alpha particles were emitted mainly from the irradiated surface of the catcher and, thus, different diagnostics were used for monitoring this emission area. However, as will be discussed also in Section 8 and 9, also some of the laser-accelerated particles from the Al target are expected to be scattered by the B surface. The use of a Thomson spectrometer may be very helpful for separating the alfa particles from the TNSA-backscattered protons. However, it has some main limitations:

i) it relies on a small pin-hole (usually not more than a few hundred micron in diameter, like in this experiment), which dramatically reduces the acceptance cone of the revealed particles, making this diagnostic tool not suitable for revealing low particle fluxes;

ii) it is unable to separate the alpha particles, produced by the fusion reactions, from the $C^{6+}$ ions, which are produced by the TNSA acceleration at the pitcher and enter the spectrometer after being backscattered by the B catcher;

iii) the spectrometer alignment with the most active part of B interacting region is definitely not an easy issue.

In Figure 5(a) it is shown an averaged spectrum for the backscattered protons achieved by directing the TS2 Thomson Spectrometer [27], designed and used for assessing ion emission and backscattering in this type of p-$^{11}$B experiments, onto the catcher. The spectrum was achieved by averaging over 71 shots, cumulated on a CR39 detector placed inside the spectrometer. The obtained spectrum is related to the trace shown on the CR39 image of Fig 5(b). No clear signature was instead obtained for other particles. This information, combined with the low energy of the detected protons, indicates that the spectrometer was potentially aligned on a region of the B catcher, distant from the centre of the main interaction area, confirming that the third limitation (mentioned above) of this diagnostic device can be quite limiting. Anyway, a clear experimental information on backscattered protons was achieved, even if not at the maximum emission point.

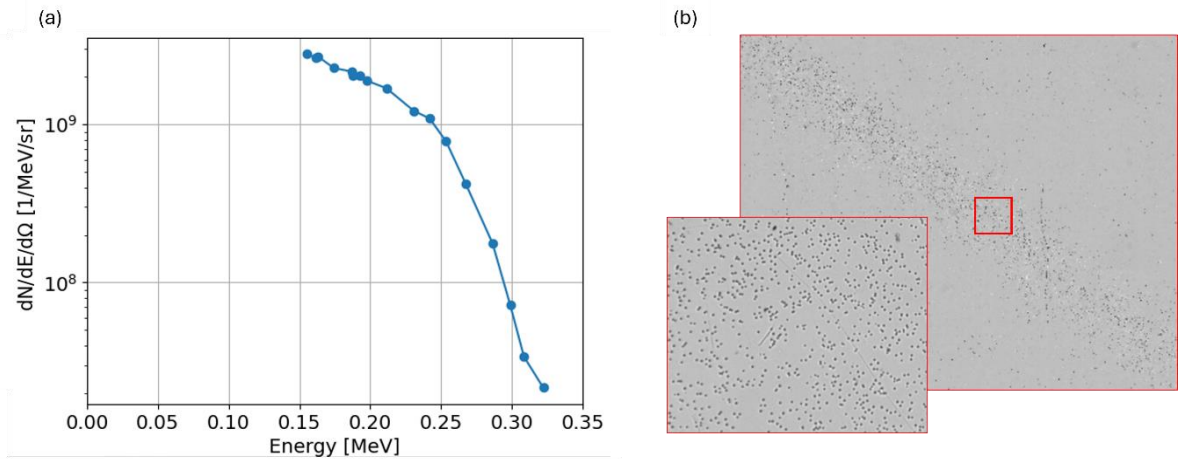

*Figure 5 - (a). Spectrum, averaged over 71 shots, of the TNSA protons backscattered from the B catcher. (b) Picture (broad and zoomed view) of the trace (associated to protons) obtained on the CR39.*

## 7. Detection of fusion products by CR 39 detectors

As discussed, the effective Boron region where alpha particles are expected to be capable of escaping is confined within about a few tens of microns from the surface. According to the previous considerations, our primary diagnostic tool for detecting escaping p-$^{11}$B fusion products consisted of the array of CR39 track detectors positioned in direct line-of-sight with the exposed B surface (see Fig 1(b)). CR39 detectors can detect single particles and have some capabilities to discern their species and energy based on the dimensions of the tracks they leave in their plastic bulk, after appropriate chemical etching [13].

To enhance detection sensitivity across different energy ranges of protons and ions, each CR39 was mounted inside a frame subdividing the exposed area in four regions, equipped each with a diverse foil acting as particle filter which, depending on the specific shot series, ranged from 2 μm PET to 40 μm of Al. Thus, the final area of each sensing region of the CR39 was of 9 x 9 mm$^2$. In Fig. 6, we show a picture of one of the detectors in our array, with the differential filters applied to the exposed regions. In Appendix C (in Table I) we provide a summary of these filter thicknesses, with the corresponding ranges for protons, alphas and carbons, obtained by Montecarlo SRIM simulations [34].

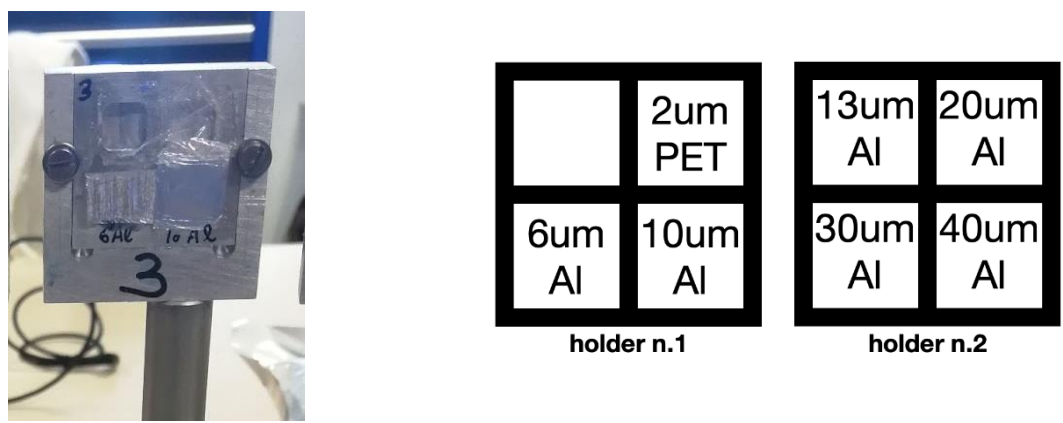

*Fig. 6 – Picture of one of the implemented CR39s, divided in four regions with different filter types (left). Complete scheme of the implemented filters (right).*

During the first part of the campaign, we evaluated the background signal on these CR39 detectors, as discussed in detail in Appendix A. In the final version of our setup, which included the thick Al shield, of key importance for allowing only particles coming from the catcher (both emitted and backscattered), we then addressed the actual capability of CR39 to discriminate protons with respect to alphas and heavier ions, which is represented by the larger tracks that the latter can produce [13], [18] on the same CR39. The calibration curves for protons and alpha particles for this set of CR39 are reported in Fig. 7 [35]. It is worth to underline that the features of the CR39s are known to change for different producers and for different sets from the same producer. Moreover, calibrations are also dependent on the type of track evaluation process (i.e. etching and microscopy) implemented. The orange dashed line on the curve of the alpha particles that is added to the calibrations for all etching times, indicates a linear response of the detector that we assume, in a conservative way, for low-energy ions.

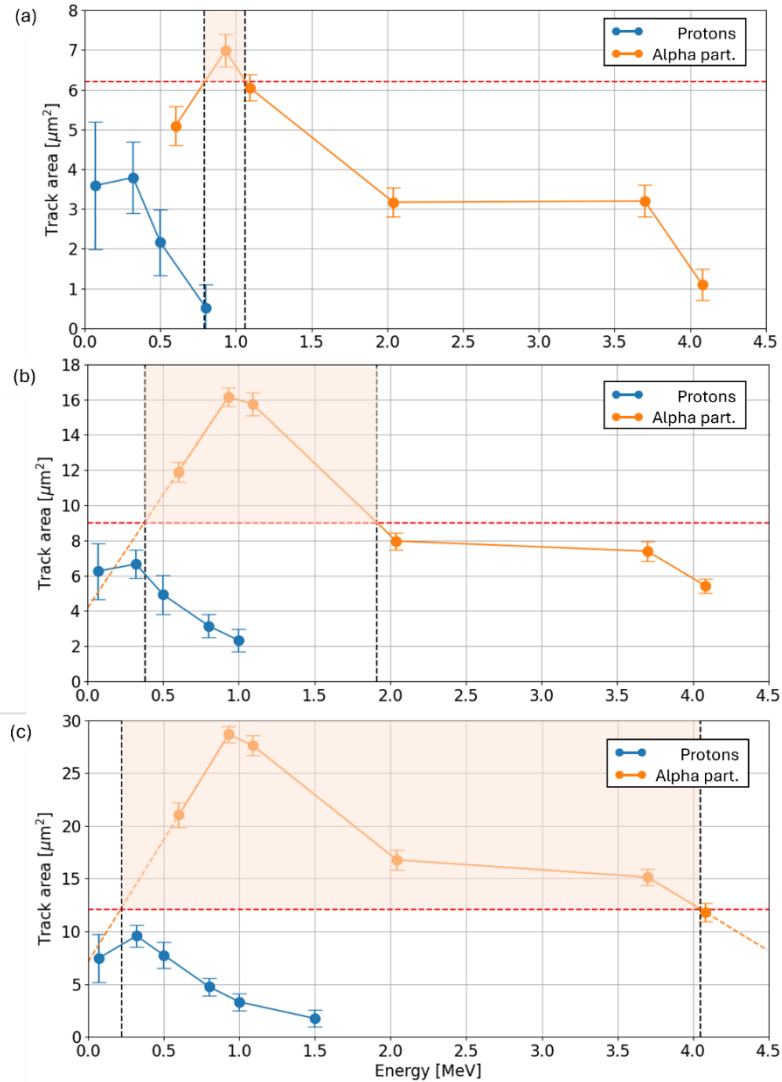

*Fig. 7 – Calibration curves of the CR39 detectors for protons and alpha particles. The reported dimensions of the traces left by the particles are for (a) 60, (b) 90 and (c) 120 minutes of etching. The red dashed lines indicate the calibration range where the track dimension is univocal, i.e. it cannot be attributed to protons. The dashed black lines indicate the corresponding energy window.*

The calibrations show the features, well known in literature [9], [13], [18], [26], [35], [36], [37], discussed in the following list. These considerations apply for both track diameter and track area, provided that circular tracks are produced, thus with no ellipticity [12].

- For each particle species, physics related to track formation and evolution during etching time gives no monotonic relationship between track area and particle energy, but rather a 'bell-shaped' curve,

- ideally starting from the (0,0) point of the coordinate axis of Fig. 7. We name here $T_p$ and $T_\alpha$ the maximum area values of these curves, for protons and alpha particles, respectively. Both depend on the etching time. It is worth to notice that calibrations at low-energies are indeed quite rare, and in general they are found in a limited energy range, due to the procedure difficulties especially in routine calibrations. In conclusion, there is no univocal correspondence of particle energy with track area.
- There is a range of areas $A$, with $A \in [0, T_p]$, where the same $A$ value finds correspondences with both protons and alpha particle curves. An example is for $A = 2$ μm$^2$ in Figure 7(a) (60 minutes etching). Therefore, we cannot discriminate protons from alpha particles in this interval of track dimension.
- In the conditions of the etching process here followed, for etching times ≥ 60 minutes, there is an energy interval of areas: $T_p < A \leq T_\alpha$ where alpha particle tracks have an area that is larger than any proton track area. We observe from Fig. 7 and from the references [9], [13], [18], [35] that the difference $\Delta_{\alpha p}(t_h) = T_\alpha - T_p$ increases with the etching time $t_h$ (at least for the examined etching conditions). These ranges are highlighted in the plots of Fig. 7 and are delimited by the dashed lines.
- For ions heavier than alphas, such as C, we can make similar considerations, since the calibrations curves have bell-shape, start from (0,0) and go asymptotically to 0 for high energies. Thus, if we define $T_c$ as the maximum value of the calibration curve for C, we may observe that $T_c > T_\alpha$ [13], [18]. So, for $A < T_c$ there is no way to separate C from alpha particles relying only on track area.

Due to these considerations, only a subset of all the tracks recorded by the CR39 detectors during the experiment can be associated to particles that are not protons. The experimental ranges where particles different from protons can be discriminated with high confidence, are those where $T_p < A \leq T_\alpha$. Their values are defined conservatively, by considering calibration uncertainties, and are indicated in Fig. 7 by the dashed horizontal red lines. For instance, for 90 minutes etching (Fig. 7(b)) the threshold of $T_p = 9$ μm$^2$ corresponds to an energy window, delimited by the black dashed lines, in the ~ [0.4, 1.9] MeV interval. Here, tracks with areal dimensions between zero and ~9 μm$^2$ can be attributed potentially to protons, alphas and any other particle such as C. These energy ranges are obtained from the calibration curves of Fig. 7 for unfiltered detectors. The use of different filters modifies the energies of the particles reaching the bare CR39 and provides, therefore, different 'energy windows' of observation to the particle beam directed towards the detector array. In Appendix C (Table 2), we indicate these 'energy windows', obtained with SRIM simulations for each used filter, for etching times of 60 and 90 minutes, used later.

In Fig. 8 we report an example of distribution of detected tracks, with respect to each filter, at position P3, in terms of particle density on the detector surface. The area of the detector without filter was saturated, and then did not allow to get a reliable distribution. The areas of the tracks were obtained after an etching procedure of 90 minutes, and show broad areal distribution, with lower values below 1 μm$^2$. Our analysis provided several information, including i) the indication of the overall number of tracks; ii) the number of tracks for which protons cannot be discriminated from heavier ions, i.e. with an area $A < T_p$, that is indicated in the plots by the blue part of the spectral curve; iii) the number of tracks due to ions different from protons (alphas, C,...), i.e. with an area $T_p < A \leq T_\alpha$, that is indicated by the red part of the spectral curve. From the latter portion of the obtained spectra, we estimated the number of potential alpha particles that impinged the CR39 detector. It is important to mention that C, N, O or other heavier ions share with alphas the capability to produce tracks larger than protons on specific energy ranges [13], [18]. These heavy ions are laser-accelerated from the pitcher Al target, have energies in the range of a few MeVs and a TNSA emission cone similar to protons. They can be scattered by the boron catcher and, like protons, reach the CR39 detectors. Thus, the actual information that we can get from these data, is the number of ions heavier than protons. We need to rely on theoretical considerations, as from the next section, to discriminate alpha particles from them, or to evaluate the percentage of alpha particles with respect to the whole set.

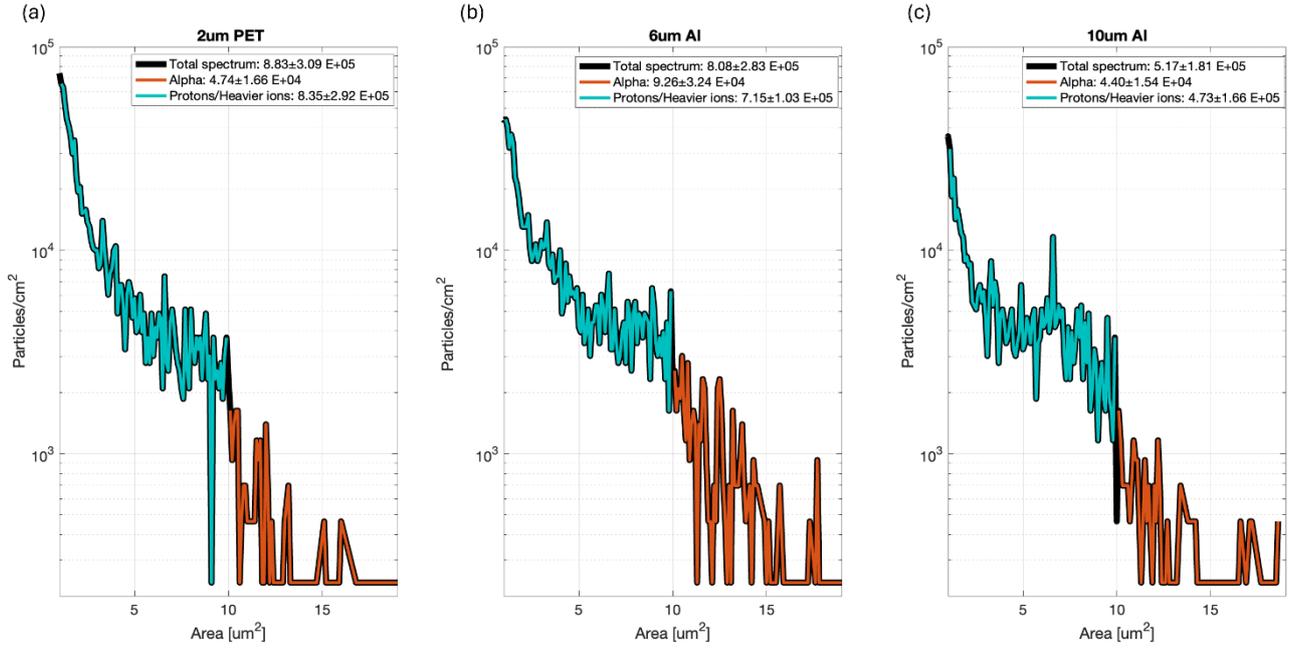

*Fig. 8 – Density of traces as a function of their areal dimensions revealed by the CR39 at position P3. The plots are for the regions with (a) 2 μm PET filter, (b) 6 μm Al filter and (c) 10 μm Al filter. The blue part of the curves indicates traces where no univocal ion specie can be attributed, i.e. they are generated by protons and/or heavy ions. The red part of the curves indicates the range where the trace dimension allows discriminating protons, based on the calibrations of Fig. 7. They are indicated in the legend as potential alpha particles, provided no heavier ions are impinging the CR39.*

### 8. Theoretical estimation of achieved fusion reactions

The use of the HPGe detector allowed investigating the nuclear activation of the proton irradiated B samples. After cumulating tens of laser-shots, the γ decays from the natural boron (composed by 80% $^{11}$B and 20% $^{10}$B) were measured, with the methodology of ref. [28]. The decay of $^{11}$C (from the reactions $^{11}$B(p,n)$^{11}$C) and $^{7}$Be (from the reactions $^{10}$B(p,α)$^{7}$Be) was detected. The γ spectrum showed a dominance of $^{11}$C decay at shorter times (due to the shorter half-life of this isotope) whereas, at longer times, only the signal from $^{7}$Be remained evident, having longer decay characteristic time. The number of reactions that were obtained experimentally is between $5 \times 10^7$ and $10^8$ for $^{11}$B(p,n)$^{11}$C and $2 \times 10^8$ for $^{10}$B(p,α)$^{7}$Be, considering the concentration of B isotopes in the irradiated sample. This provides a robust feedback for the analytical model that was proposed in Ref. [28], which allows calculating an estimation of the number of induced $^{11}$B(p,α)$^{8}$Be reactions, resulting in $2.2 \times 10^8$ overall fusion events of this type. Using the proton beam measured by TS 1 (see Fig. 2(a)) and assuming that it is almost entirely directed onto the B catcher (as also indicated by the microscope image of Fig. 4(a)), the analytical model provides the expected spectrum of the produced alpha particles, which is reported in Fig. 9. This spectrum (red plot) represents the fraction of alpha particles that escape the catcher and reach the location of the CR39 detectors, among those that are produced in the bulk or on the surface of the B catcher, having sufficient initial energy to exit the B bulk, according to their point of origin inside it. They were calculated considering the penetration depth of the protons driving the reaction. In particular, the number and energy of the alpha particles were estimated for each proton energy. The energy loss of alphas in the B catcher, and later in different filters that were used in the experimental CR39 configurations, were estimated using SRIM. Different alpha channels were estimated using the experimental cross sections for both the $^{10}$B and $^{11}$B components. While the $^{10}$B case is a simple two-body kinematics, alphas produced from the $^{11}$B were estimated assuming also a two-body kinematics with $^{8}$Be. We followed the approach of Kimura et al. [8] and assumed a main contribution for this channel, denominated α$_1$, where $^{8}$Be is in its first excited level. A Breit-Wigner with the experimental width of 1.5 MeV was used to randomly generate the alpha and $^{8}$Be* excited state kinetic energies. The $^{8}$Be* fragment subsequently decays into two alphas. The particles which are emitted towards the direction of the CR39, and energetic enough to pass through the filters, are eventually detected. The first peak, at around 2.5 MeV energy observed in the theoretical distribution of Fig.9, is due to the decay of $^8$Be* from the excited level. More details will be given in a follow up paper.

In the same figure, with the blue curve, we report the spectrum of protons scattered by the surface of the B catcher and that are deviated towards the angular position of the CR39 detectors, i.e. with an angle of ∼ 54° with respect to the B surface (see Fig1(b)), according to Rutherford's scattering model [38]. Concerning C ions, which have mass slightly larger than B, Rutherford's scattering formulas give a maximum deflection angle of ∼ 26° from the B surface. Therefore, they cannot reach the CR39 array (as one can see from Fig. 1(b), remembering that distances are not in scale to each other). However, if the catcher atoms had a mass slightly larger than B, the scattering angles would change significantly. This is the case, for example, for BN catchers (often used in p-$^{11}$B experiments [11], [30], [39]) or, alternatively, in the case of impurities deposited on the B surface, which alter the effective atomic mass of the scattering surface material. The latter effect is one of the typical drawbacks of accumulating numerous shots on the same B catcher and requires a detailed analysis of the deposited materials and the thickness of the deposition, which is out of the purpose of the present paper. Instead, we consider here the simple case of a BN catcher, with effective ion mass of A=12 like the incoming C ions. This represents a pejorative scenario, compared to our experimental conditions, overestimating the number of scattered C ions. The results are shown in Fig. 9 orange curve. In Fig. 9(b), we report of the low energy part of these spectra, for the two cases of 2 μm PET and 10 μm Al filters, which were some of those implemented on our CR39 detectors. The particle flux, especially for the C ions, as expected, decreases compared to the filter-less case. For the 10 μm Al case, due to the higher stopping power and thickness, the orange curve of the C ions is decreased by about two orders of magnitude. The distribution of alpha particles (red curves), due to the deeper penetration depth remains, suffers much less decrease. This analysis allows us to evaluate the contribution of scattered carbon ions from BN on the CR39s, which is important, since their signals (i.e. track dimensions) are similar compared to the alpha particles [18]. Even though overestimating the number of scattered C ions, as mentioned before, we obtained that their contribution does not influence the measurements obtained by the CR39s significantly. If we consider the detected alpha particle energy range in the typical case of 90 min etching, i.e. from 0.38 to 1.91 MeV, only in in the case of the 2 μm PET filter (i.e. the thinnest filter of our set, representing only the first point of the spectral curve that we will discuss in the following section), the number of C ions is comparable to the number of predicted alpha particles. For this filter thickness the contribution of alpha particles and C ions on the detector is estimated to be of 40% and 60%, respectively, which might, therefore, lead to an overestimation of the obtained alpha particle flux. In the case of the thicker filters, such as for the case of 10 μm Al filter shown in panel (b), the contribution of carbon ions is more than one order of magnitude lower than the number of expected alphas and, therefore, negligible.

The primary number of alpha particles, predicted by this model, that shall be detected by the detectors, compares very well, with those experimentally revealed in the CR39, as will be discussed in the following section. A similar agreement is found for the production of $^{11}$C and $^7$Be as discussed in Ref. [28]. The model agrees quite well with the measurements obtained for scattered protons measured from the Thomson Spectrometer TS2 (Fig. 6(a)) for energies higher than about 300 keV. For lower energies the model might be underestimating the scattering of low-energy protons from B, since they are not measured by the TS1, because of intrinsic lower threshold.

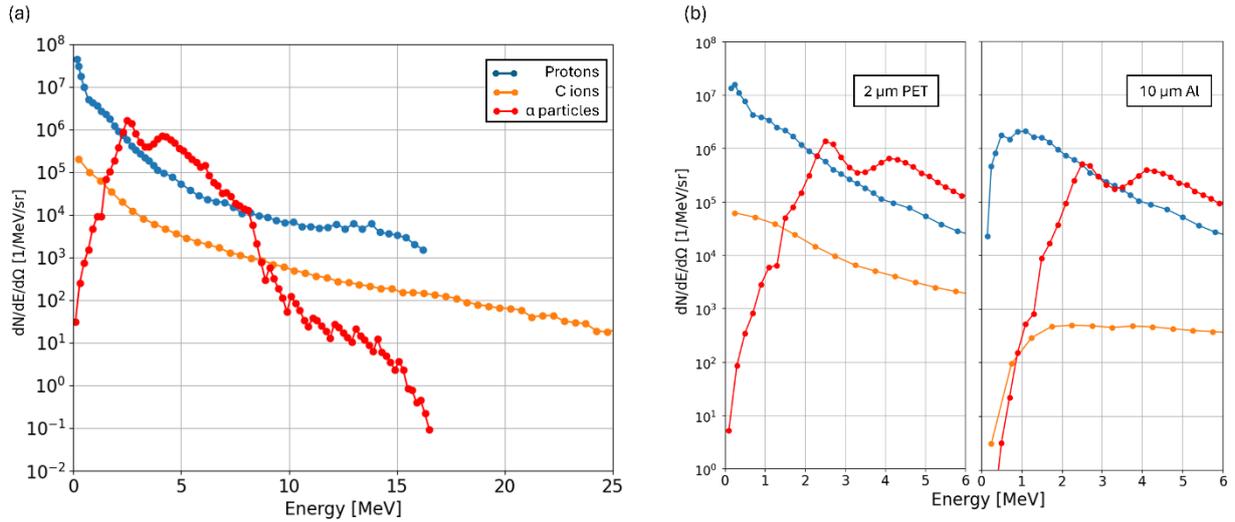

*Fig. 9 – (a) Spectra of scattered protons (blue), scattered carbon ions (orange), in the case of equal atomic mass in catcher (orange) and alpha particles that reach the position of the CR39 detector (red). In this case, no filter is considered, i.e. the spectra are calculated considering the particles as they are scattered/generated at the B catcher. (b) Spectra of scattered protons (blue), scattered carbon ions (orange) and alpha particles that pass through the filters that were used in the CR39 experimental configurations (red).*

## 9. <u>**Spectral analysis of the measured alpha particles**</u>

By performing the discrimination of the protons, using the method of Section 7, for the cases of different filters on the CR39s, we obtained a spectral information about the distribution of particles, different from protons, that impinge our detectors. In Fig. 10 we report data by integrating the number of track counts of the red part of the obtained plots of track density, like the example of Fig. 8. The different energy windows, indicated by the horizontal bars around each point, were obtained by adapting the highlighted energy ranges of the calibration curves of Fig. 7, to the filter thicknesses used in front of the detectors (see Appendix C, Table II). We then used these numbers for estimating the original spectrum of the incoming particles, before to pass through the filters. In order to do so, we made the assumption that all those particles were alpha particles, and then performed SRIM simulations for accounting of energy loss on the several filters, and got the initial energy of the alpha particles when they left the catcher. The values of particle number reported in Fig. 10 are obtained by normalizing the track density (expressed in number of tracks per unit of area on the detector) to the solid angle intercepted by the exposed area of the CR39s and normalized by the width of the energy window they refer to. They are then normalized to the number of cumulated shots. In this way, we are capable to supply experimental estimation of the spectrum of alpha particles reaching the CR39 array. The strict integration of these curves gave the total number in these specific energy intervals, obtaining $\sim 2.1 \times 10^6$ alpha/sr for the first shot series (black dots) and $\sim 1.5 \times 10^6$ alpha/sr for the second series (green dots). The values indicated by the red diamond markers have been obtained from the analytical model of Section 8. We integrated the analytical curve of the generated alpha particles (Fig. 9(a), red curve), which considers the particle flux that escapes the B catcher and impinges the CR39 detectors, within the energy ranges of the experimentally obtained points of the black dataset. The values of particle number predicted by the model are indicated by the red dots and, by integration of these values, the total number of predicted alpha particles in the range from 1.6 MeV to 5.4 MeV is of $\sim 2.2 \times 10^6$ alpha/sr.

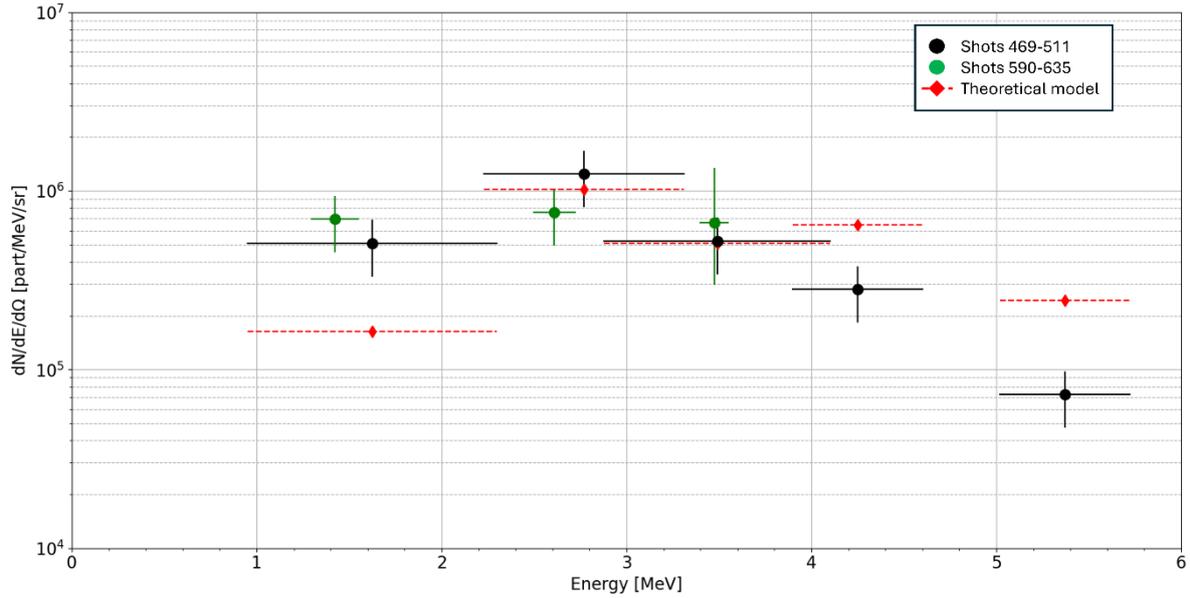

*Fig. 10 - Estimated alpha spectrum detected by the CR39 array. The black dots were obtained with an etching time of 90 minutes, for a 43 shot series. The green dots were obtained with a 60 minutes etching time for a 46 shot series. The red curve represents the analytical spectrum of expected alpha particles that impinge the detector.*

10. **Discussion**

The purpose of this work was to assess the features and potential of the pitcher – catcher configuration for driving p-$^{11}$B nuclear reaction with PW high-repetition-rate lasers. For this reason, we gave particular care to the characterization of the interaction and to the accelerated/scattered particles. Our findings confirm that this scheme is suitable for producing alpha particles using PW class, high-repetition-rate laser system. We therefore discussed here a methodology for effectively evaluating the performance of such laser-driven alpha source and addressed the challenges of implementing an experimental setup of this type. The proposed methods, which can be adapted to experiments that make use of similar setups, essentially rely on i) careful characterization of the TNSA beam (addressing both protons and heavy ions) that drive the fusion reactions; ii) analytical/numerical modelling and/or additional measurements that allow estimating the flux of the produced alpha particles that escape the B catcher, with respect to flux of protons and heavy ions that potentially produce a similar signal on the implemented detectors; iii) analysis of the signals obtained from CR39 detectors that allows excluding the contribution of protons from heavier ions, which produce tracks compatible with alpha particles.

The TNSA beam was monitored by multiple diagnostics for obtaining a full characterization of the accelerated particles. The measurements, routinely performed during the entire campaign, allowed a tuning of the laser parameters for an optimal interaction with the Al target. Moreover, exploiting the high repetition rate of the laser, they allowed obtaining typical proton and carbon spectra from a statistical analysis of tens of dedicated shots. These spectra provided therefore a statistically sound input for the analytical model that aimed at predicting the flux of produced alpha particles.

A first, clear information that a significant amount of p-$^{11}$B → 3α fusions occurred in the B was provided from the measurements of activation by the HPGe detector. As discussed in Section 8, an estimated overall number of alpha particles produced in the B catcher was obtained, equal to $6.6 \times 10^8$ α/shot. However, as it was also observed from the experimental data, due to the broad spectrum and high energy of the proton beam impinging the B, the alpha particles are produced in the whole bulk of the B catcher. Only protons with a few MeV energy at most, trigger fusion reactions generating alpha particles close enough to the surface, that escape from the front side of the boron bulk. Thus, our analytical model predicts an alpha spectrum that is strongly modified with respect to the original one, produced within the bulk. This is described in Section 8, where also the effect of different filter thicknesses on the particle spectra (alpha particles and scattered particles) is analysed. This analytical approach allowed us interpreting the results obtained from the detectors used for revealing the alpha particles, since our observations indicated that they were indeed produced by both p-$^{11}$B fusion reaction products and particles scattered by the B sample.

The main diagnostic tool for revealing the alpha particles stemming out of the B catcher was an array of CR39 placed in line of sight with the irradiated B surface. During the first part of the campaign, we optimized the position of these detectors and routinely performed preliminary analysis on-site. This allowed estimating the maximum number of shots to be cumulated on the detector before saturating it, i.e. between 35 and 50 shots. Moreover, we performed a careful estimation of the background signal that was produced on the detectors, leading to the implementation of the protecting Al shield (see Fig. 1(b)), which allowed removing a significant background signal generated by the direct irradiation of the TNSA beam (see Appendix B). As discussed in Section 7, the calibration curves of the used CR39 detectors can be used for obtaining threshold values of the track dimensions that allow discriminating the contribution of protons. This approach, in combination with equipping the CR39 array with differential filtering and a time-step etching, can give a suitable estimation of the impinging alpha spectrum, if information on particles heavier than protons that reach the array (such as TNSA particles scattered from B) are supplied. As obtained from Figure 9, in the alpha particle energy range of interest, from 0.38 MeV to 1.91 MeV, in the case of 90 min etching of the CR39 and a 2 μm PET filter in front of the detector, the number of estimated C ions is comparable to the predicted number of alpha particles, for the example case of BN catcher. A possible overestimation of the alpha flux due to the contribution of C ions is limited to this only case, since the use of thicker filters to protect the detector, leads to significantly lower contribution of carbon ions, down to more than one order of magnitude, in the case of 10 μm (or thicker) Al filter. This shows that the C ion contribution is negligible in the spectrum of Figure 10, representing the alpha particles reaching the CR39 array. The total flux of measured alpha particles, considering the black dataset of Fig. 10 (i.e. the one over the widest range of energy), obtained after a 90 minutes etching procedure, is estimated ~2.1±0.7 × 10$^6$ alpha/sr, which compares very well with the theoretical flux of alphas (~2.2 × 10$^6$ alpha/sr) calculated from red dots of Figure 10 in the same energy range. Considering a uniform alpha emission from the overall number of estimated fusion reactions, obtained from HPGe measurements, it is possible to have a reference number for the estimated alpha density flux originating within the B bulk over the whole alpha spectrum: 5.25 × 10$^7$ alpha/sr. The number of alpha particles detected by the CR39 in the limited energy range of the black dataset of Fig. 10, are about 4% of the latter, which are instead related to the full spectral emission. This gives us an estimation of the minimum transmission ratio of alphas produced in the B bulk, which are capable to pass through it and reach the CR39 position in this experiment. The shape of the theoretically expected alpha particle distribution and, consequently the obtained red dataset of Fig. 10, although predicting a comparable integrated particle flux, have some differences from the experimental data. The reasons are likely to be the following: i) the theoretical model calculates the alpha particle spectrum, starting from the measured proton energy spectrum, which (as shown in Fig. 2) has significant fluctuations; ii) the analytical curve is based on models (concerning, e.g., the cross-sections for the fusion events and the penetration range of the accelerated protons and generated alpha particles) that assume the B catcher to be solid and at room temperature; the effects due to impinging X-rays, hot electrons (Fig. 3)) and high energy protons have on the B surface, are not considered (heating and plasma formation, damaging the catcher surface within a large sequence consecutive shots and producing ion implantation on catcher surface,…). Future perspectives of this work include the development of an even more accurate theoretical model for predicting the experimental measurements in a real-case scenario of high-repetition-rate laser-matter experiments.

The work was capable to underline, as also shown in the Appendixes, which are the issues to be taken into account for a suitable use of this setup for the purpose described, which were then effectively implemented in later experiments of the same type on the same facility [30], [39] . The scheme can be thus effectively applied to high repetition rate petawatt scale laser facilities with higher energy and intensity in order to exploit its full potential. An experimental campaign of this type has been recently performed at ELI-Beamlines with L3 laser by part of this group and the results are now under preliminary assessment.

**Appendix A: Etching procedure of the irradiated CR39 detectors**

After being exposed to the particles and the radiation produced during the laser-plasma interactions, the detectors were immersed in an aqueous solution of 6.25 mol/L of sodium hydroxide (NaOH) at 70 ± 0.1 °C for a variable etching time. This process aimed at providing comprehensive insight into the evolution of track diameter over etching time, thereby enhancing the potential for particle discrimination, at expenses of increased throughput of data. After the etching process, the CR39 detectors were analysed to extract valuable data from the visible tracks, using the Nikon ECLIPSE Ni-E fully automated inverted optical microscope, enabling precise and detailed track measurements. Pictures were acquired with a DS-Qi 2 camera, having a resolution of 4908x326 and a sensitivity range of ISO 800-51200 mV/s.

**Appendix B: Background signal on the irradiated CR39 detectors**

In the present campaign, the CR39 detectors revealed particles stemming from the catcher, over tens of laser shots. Given the expected low number of alpha particles, during the first part of the campaign special attention was paid to estimating the background on the CR39 detectors without catcher and related holder. This estimation considered two main aspects:

1. latent tracks due to natural background on the CR39, caused by environmental radiation [13];
2. particles reaching the detectors due to scattering off the surfaces of the TNSA target frame and of the various objects in the chamber.

Concerning the first point, the frame where the detectors were placed was made of 2 mm-thick Al, capable to be an effective shield for any of the particles meant to reach the CR39. So, the etching of the not-exposed regions of the CR39 supplied suitable information on the background, that resulted very low for this brand-new set of CR39s: 10 tracks/cm$^2$.

We estimated the second type of contribution by analysing some CR39s in the absence of both the catcher and its holder. We found that a surprisingly large number of ions was able to reach them. In Fig. 11 (top picture), we report the image from one exposed region of a CR39 placed in position (P2), obtained after a series of 86 shots and chemical etching. It is evident that the region of the CR39 is heavily polluted by particles with different areal dimensions. In order to overcome this problem, we found that the careful positioning of a protecting Al screen (shown in Fig.2) was an effective solution. In Fig 11 (bottom picture) we show an image of a CR39 etched after 20 shots with the Al screen implemented. It is apparent that the use of the thick Al plate achieved no visible pollution of the detector) and then this served well at stopping all the stray particles observed in the top picture. This showed that one significant contribution to that pollution comes from the pitcher structure and sets an important care that must be followed for future implementations of this pitcher-catcher scheme for p-$^{11}$B interaction

To a more careful observation, the thick pitcher frame was built with a conical hole on the back side of 45° half aperture angle, and the CR39s were placed at about 90 degrees angle from the normal axis of the pitcher (see Fig 1(b)). So, no direct line of sight was provided from the latter to the three CR39s. Nevertheless, potential lateral straggling of ion flows, interacting with the frame edges, should have produced the important scattering component giving stray ions reaching the CR39s. As a matter of fact, the use of the Al shield was effective on protecting the detectors at P2 and P3 positions of the CR39 array (see Fig 1(b)) and partially mitigated the pollution on the detector on P1. These observations showed that ion scattering was one issue to take into important care, and for this reason we wanted to minimize potential contributions that may have arisen from scattering of the TNSA ions emitted from the pitcher on the B frame holder, by considering in the analysis of the CR39 images the selection of tracks with no or very small (0.9) ellipticity, so coming from the B surface and its close neighbourhoods, rather than potential other objects nearby [12].

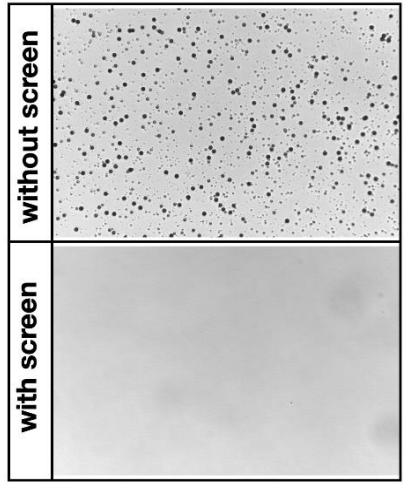

*Fig. 11 – CR39 microscope images for the pitcher-catcher setup, with and without the Al shield protecting the detector array. The top image indicates the case without the protecting Al screen, after a series of 86 shots, without the boron catcher. In the bottom row, a CR39 at the same position, after a series of 20 shots without the catcher, with the protecting Al screen in place.*

**Appendix C: Characterization of the differential filtering used in the CR39 array**

In Table I, we report the cut-off energies of protons and heavier ions provided by the different filters that were used for covering the exposed regions of the CR39 detectors.

In Table 2, we report the energy ranges of alpha particles, that meet the condition of tracks on the CR39 detector with an area $T_p < A \leq T_\alpha$ according to calibrations [35]. These ranges are listed for the different filter thicknesses and materials used in the experiment. The range in the case of no filter, is obtained from the calibration curves of Fig. 7. The ranges for the filtered regions, are obtained by SRIM simulations.

*Table I. Cut off energies for each used filter*

| Filter thickness [μm] - material | Cut-off energy for protons [MeV] | Cut-off energy for alpha particles [MeV] | Cut-off energy for C [MeV] |
|---|---|---|---|
| 2 - PET | 0.17 | 0.35 | 1.05 |
| 6 - Al | 0.53 | 1.73 | 6.63 |
| 10 - Al | 0.77 | 2.74 | 11.63 |
| 13 - Al | 0.93 | 3.45 | 20.45 |
| 20 - Al | 1.23 | 4.70 | 22.18 |
| 30 - Al | 1.61 | 6.26 | 30.80 |
| 40 - Al | 1.93 | 7.60 | 38.28 |

*Table II. Results, obtained by SRIM simulations, about the thresholds associated with Fig. 7 for the several etching times, and the associated correspondence on incoming alpha particles energies when different filters are used.*

| Filter thickness [μm] - material | 60 minutes etching - Alpha particles threshold energies [MeV] | | | | 90 minutes etching - Alpha particles threshold energies [MeV] | | | |
|---|---|---|---|---|---|---|---|---|
| | *E min.* | *E max.* | *ΔE* | *E mean* | *E min.* | *E max.* | *ΔE* | *E mean* |
| *No filter* | *0.79* | *1.06* | *0.27* | *0.92* | *0.38* | *1.91* | *1.53* | *1.14* |
| *2 - PET* | *1.3* | *1.55* | *0.25* | *1.42* | *0.95* | *2.3* | *1.35* | *1.62* |
| *6 - Al* | *2.5* | *2.72* | *0.22* | *2.61* | *2.23* | *3.31* | *1.08* | *2.77* |
| *10 - Al* | *3.4* | *3.55* | *0.15* | *3.47* | *2.88* | *4.1* | *1.22* | *3.49* |
| *13 - Al* | *3.97* | *4.12* | *0.15* | *4.04* | *3.9* | *4.6* | *0.7* | *4.25* |
| *20 - Al* | *5.15* | *5.3* | *0.15* | *5.22* | *5.02* | *5.72* | *0.7* | *5.37* |

**Acknowledgements**

This work has been carried out within the framework of the EUROfusion Consortium, funded by the European Union via the Euratom Research and Training Programme (Grant Agreement No 101052200 — EUROfusion). Views and opinions expressed are however those of the author(s) only and do not necessarily reflect those of the European Union or the European Commission. Neither the European Union nor the European Commission can be held responsible for them. The involved teams have operated within the framework of the Enabling Research Project: ENR-IFE.01.CEA "Advancing shock ignition for direct-drive inertial fusion".

This work has been carried out within the framework of the COST Action CA21128- PROBONO "PROton BOron Nuclear fusion: from energy production to medical applications", supported by COST (European Cooperation in Science and Technology - www.cost.eu)

The research leading to these results has received funding from LASERLAB-EUROPE (grant agreement no. 871124, European Union's Horizon 2020 research and innovation programme).

**Data availability**

The data that support the findings of this study are available from the corresponding authors upon reasonable request.